\documentclass[11pt,twoside]{article}

\usepackage{baaa-engb2}

\usepackage{graphicx}

\usepackage[T1]{fontenc} % Computer Modern (CM) fonts

\usepackage{latexsym}
\usepackage{verbatim}

\begin{document}
\def\tablename{Tabla}%

%\vskip 1.0cm
\markboth{T.A. Michtchenko, S. Ferraz-Mello \& C. Beaug\'e}%
{The periodic and chaotic regimes of motion in the exoplanet 2/1 mean-motion resonance}

\pagestyle{myheadings}
%
%%%% UNCOMMENT THE LINE CORRESPONDING TO YOUR KIND OF CONTRIBUTION %%%%%
%
\vspace*{0.5cm}
%\parindent 0pt{LECTURE}
%\parindent 0pt{ORAL COMMUNICATION }

%\vskip 0.3cm
\title{The periodic and chaotic regimes of motion in the exoplanet 2/1 mean-motion resonance
%(Template of article for the contributions of the
%\textit{$3^{rd}$ La Plata International School on Astronomy and Geophysics})
}

\author{T.A. Michtchenko$^{1}$, S. Ferraz-Mello$^{1}$, C. Beaug\'e$^{2}$}

\affil{%
  (1) Instituto de Astronomia, Geof\'isica e Ci\^encias Atmosf\'ericas, USP, S\~ao Paulo, Brazil\\
  (2) Observatorio Astron\'omico, Universidad Nacional de C\'ordoba, C\'ordoba, Argentina
}

\begin{abstract}
We present the dynamical structure of the phase space of the planar planetary 2/1 mean-motion resonance (MMR). Inside
the resonant domain, there exist two families of periodic orbits, one associated to the librational motion of the critical angle ($\sigma$-family) and the other related to the circulatory motion of the angle between the pericentres ($\Delta\varpi$-family). The well-known apsidal corotation resonances (ACR) appear at the intersections of these families. A complex web of secondary resonances exists also for low eccentricities, whose strengths and positions are dependent on the individual masses and spatial scale of the system.

Depending on initial conditions, a resonant system is found in one of the two topologically different states, referred to as  \textit{internal} and \textit{external} resonances. The internal resonance is characterized by symmetric ACR and its resonant angle is $2\,\lambda_2-\lambda_1-\varpi_1$, where $\lambda_i$ and $\varpi_i$ stand for the planetary mean longitudes and longitudes of pericentre, respectively. In contrast, the external resonance is characterized by asymmetric ACR and the resonant angle is $2\,\lambda_2-\lambda_1-\varpi_2$. We show that systems with more massive outer planets always envolve inside internal resonances. The limit case is the well-known asteroidal resonances with Jupiter. At variance, systems with more massive inner planets may evolve in either internal or external resonances; the internal resonances are typical for low-to-moderate eccentricity configurations, whereas the external ones for high eccentricity configurations of the systems. In the limit case, analogous to Kuiper belt objects in resonances with Neptune, the systems are always in the external resonances characterized by asymmetric equilibria.
\end{abstract}

\section{Introduction}\label{sec:0}

Asteroids and exoplanets are amongst the most striking sources of problems in Celestial Mechanics where chaos and order play an important role. Asteroids have always puzzled astronomers by their peculiar distribution: in some regions, they avoid resonances, but in others, they prefer to be exactly where the resonances are found. The famous problem of the Kirkwood gaps, solved in the last quarter of the past century, is the better example. Asteroids have benefited from more than one century accurate observations and the number of asteroids whose orbits are known with high precision is presently of order of many thousands. Accurate observations allowed us to map the phase space where the asteroids evolve and obtain the boundaries of regular and chaotic motions. In addition, the increasing computer capacity allowed us to understand why they are distributed in such a way.

Exoplanets on the other hand are being observed at the limit of our technical capabilities. We are far from having accurate knowledge of their orbits. We know presently more than 700 planets. Most of them are in systems composed of only one planet. But the number of multi-planet systems is increasing and may reach its first hundred in a short time. Besides, what is important, they show a striking diversity. They range from the small systems of super-Earths and mini-Neptunes, as those discovered by the space missions CoRoT and Kepler in circular orbits very close to their central stars, to systems of giant planets in very elongated elliptic orbits reaching enormous distances from
their host stars, as far as the outskirts of our Solar System. Eccentricities larger than 0.5 are not infrequent: in multi-planet systems they reach up to 0.75 and in single-planet systems they reach up to 0.97.

Many multi-planet extrasolar systems exhibit resonant behavior. Systems of resonant planets is a novelty for the Celestial Mechanician. We were used to study the resonant problems of small objects, whose mass can be assumed as zero (restricted three-body problem), or a few planetary satellites in which restricted models cannot be used but which are dominated by the oblateness of the central planet. So, exoplanets are a prime source for cases of the three-body problem in which we cannot neglect the mass of the bodies and in which the mutual point-like gravitational interaction is by far the dominant force. The most frequent resonance found is the 2/1 mean-motion resonance (e.g. GJ\,876 \textbf{c}-\textbf{b}, HD\,40307 \textbf{c}-\textbf{d}, HD\,73526 \textbf{b}-\textbf{c}, HD\,82943 \textbf{c}-\textbf{b} and HD\,128311 \textbf{b}-\textbf{c}).

Although other resonance sites may also be inhabited (e.g. HD\,60532 \textbf{b}-\textbf{c} in the 3/1 or HD\,45364  \textbf{b}-\textbf{c} in the 3/2 resonance), the 2/1 is presently the most populated. Notwithstanding the limited accuracy of the observations, resonant configurations have at least two strong arguments in their favor: The long-term stable motion of close planets in high eccentricity configurations is possible only if the planets are locked in (and protected by) MMRs. Secondly, resonance trapping appears to be a natural outcome of planetary migration processes due to planet--disk interactions, which are believed to take place in the latest stage of the planet system formation (e.g. Kley 2000; Snellgrove et al. 2001; Kley et al. 2005).

The classical studies of the general three-body problem with their averaged Hamiltonians have been quickly extended to the case of exoplanets (Hadjidemetriou 2002; Lee \& Peale 2002; Beaug\'e \& Michtchenko 2003; Ferraz-Mello et al. 2003). The periodic solutions, dubbed as corotation resonances or apsidal corotation resonances (ACR), were obtained  to cover all possible families in low-order resonances (Lee 2004; Beaug\'e et al. 2006; Michtchenko et al. 2006b, Giuppone et al. 2010). ACR are stationary states of the resonant Hamiltonian averaged over the synodic period and correspond to equilibrium solutions of the averaged equations of motion.

Notwithstanding the attention devoted to ACR and periodic orbits, not much was known on the topology of the phase space outside their vicinity. A detailed analysis was presented first by Michtchenko \& Ferraz-Mello (2001) for the 5/2 MMR, by Callegari et al. (2004) for the 2/1 MMR and by Callegari et al. (2006) for the 3/2 MMR. However, these works were limited to the study of a few specific systems and are only valid for very small eccentricities. Even so, results showed the complex structure of the phase space, populated by several different families of periodic orbits, modes of oscillation and possible regimes of motion. ACR appear to be only one of several distinct types of stable configurations.

It was necessary to extend our knowledge of the resonant dynamics of the MMR beyond the ACR. This has been done for the 2/1 resonance in two papers published by Michtchenko et al. (2008 a,b), for arbitrary mass ratios and with no restrictions in the orbital eccentricities, providing a complete map of the phase space around the ACRs. These results, summarized in this paper, help us understand the results of numerical simulations and pinpoint where future (or even currently known) extrasolar systems may be found. Even if all planetary systems in this commensurability seem to be restricted to the very close vicinity of ACR, the maps allow to see the possible evolutionary routes inside the 2/1 MMR from initially non-resonant configurations.

\section{The model of the resonant three-body problem}\label{sec:1}

When the ratio of the orbital periods of two planets is close to a ratio of two simple integers, we say that the planets are close to a mean-motion resonance (MMR). A mean-motion resonance is often written in a generic form as
$$
n_1/n_2=(p+q)/q,
$$
where $n_i$ are the planetary mean motions and $p$ and $q$ are integers, the latter one representing the order of the resonance. The critical (or resonant) angles $\sigma_i$ are defined as
\begin{equation}
\begin{array}{ccl}
\sigma_1 &=& (1+p/q) \lambda_2 - (p/q) \lambda_1 - \varpi_1, \\
\sigma_2 &=& (1+p/q) \lambda_2 - (p/q) \lambda_1 - \varpi_2,
\label{eq2}
\end{array}
\end{equation}
where $\lambda_i$ and $\varpi_i$ are mean longitudes and longitudes of pericentre of the planets; index $1$ denotes the inner body, while index $2$ is reserved for the outer body. The behavior of the critical angles define the location of the system with respect to the resonance: when one of these angles is librating, the system is said to be {\it inside} the resonance. It is worth noting that the secular angle defined as a difference in longitudes of pericentre of the planets, is $\Delta \varpi = \sigma_1-\sigma_2$.

The dynamics of two resonant planets, with masses $m_1$ and $m_2$, orbiting a star of the mass $m_0$ is defined by the
averaged Hamiltonian ${\overline {\mathcal H}_{\rm res}}$ and two integrals of motion, ${\mathcal AM}$ and ${\mathcal K}$, where the first is the total angular momentum and the second is the so-called spacing parameter. These
analytical functions are given by the expressions, up to second order in masses:
\begin{equation}
\begin{array}{ccl}
{\overline {\mathcal H}_{\rm res}} &=& -\sum_{i=1}^2{\frac{Gm_0m_i}{2\,a_i}}-\frac{1}{2\pi}\int_0^{2\pi}\,
{\mathcal R}(a_i,e_i,\sigma_i,Q)\,dQ, \\
             & & \\
{\mathcal AM} &=& m_1\,n_1\,a_1^2\,\sqrt{1-e_1^2}+m_2\,n_2\,a_2^2\,\sqrt{1-e_2^2} {\rm ,} \\
             & & \\
{\mathcal K} &=&(p+q)\,m_1\,n_1\,a_1^2+p\,m_2\,n_2\,a_2^2{\rm ,}
\label{eq1}
\end{array}
\end{equation}
where ${\mathcal R}$ is the disturbing function. All orbital elements, including the semi-major axes $a_i$ and eccentricities $e_i$, are canonical astrocentric variables; their relation to the usual osculating heliocentric orbital elements can be found in (Ferraz-Mello et al. 2006). The averaging of the disturbing function ${\mathcal R}$ is done with respect to the synodic angle $Q=\lambda_2-\lambda_1$. In the vicinity of the MMR, $Q$ is much faster than the resonant and secular angles, and does not influence significantly the long-term evolution of the system. Thus, all periodic terms dependent on $Q$ can be eliminated (i.e. averaged out) of the Hamiltonian function, and only secular and resonant terms need to be retained.
\begin{figure}[!ht]
  \centering
  \includegraphics[width=.45\textwidth]{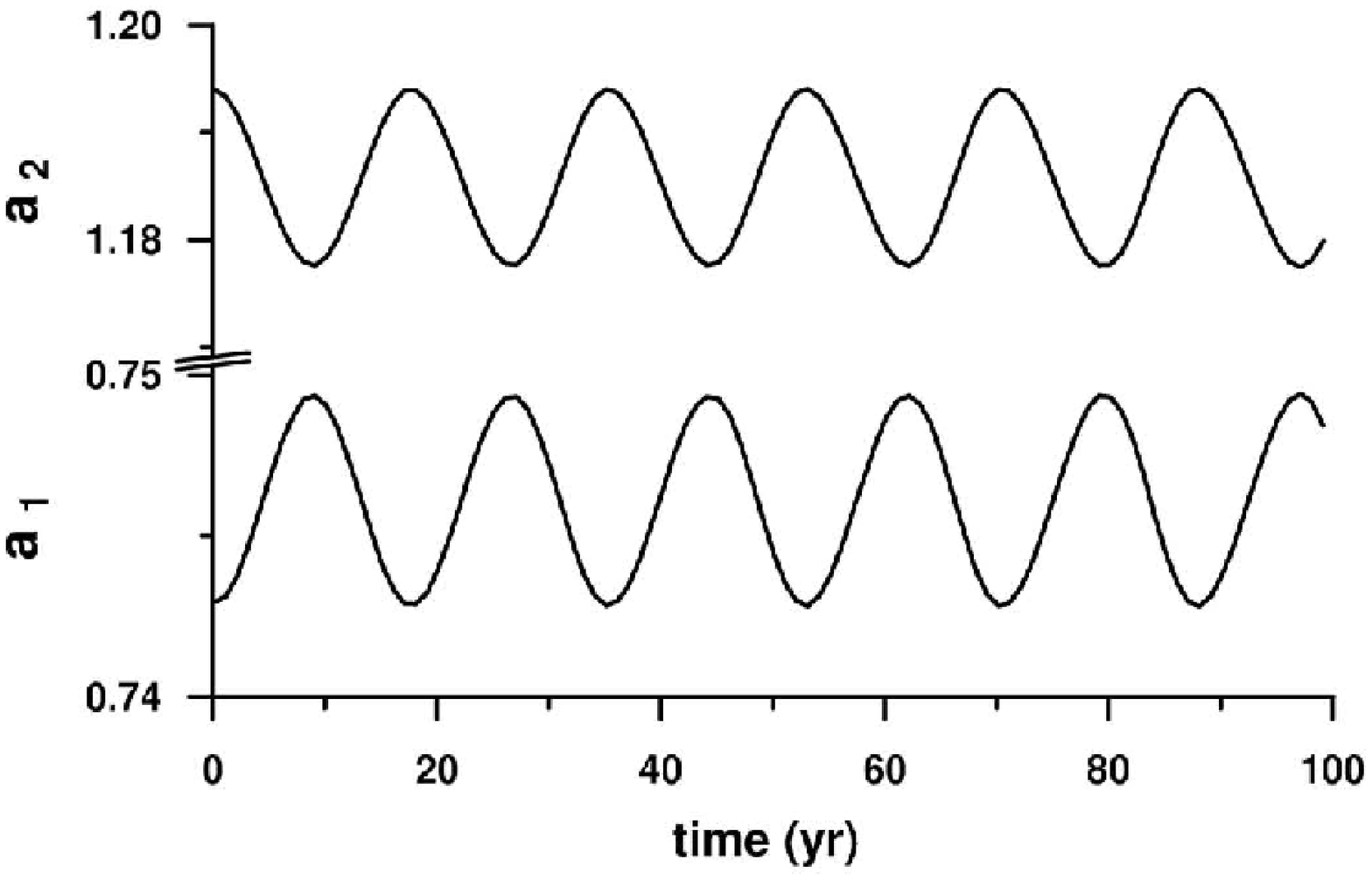}~\hfill
  \includegraphics[width=.45\textwidth]{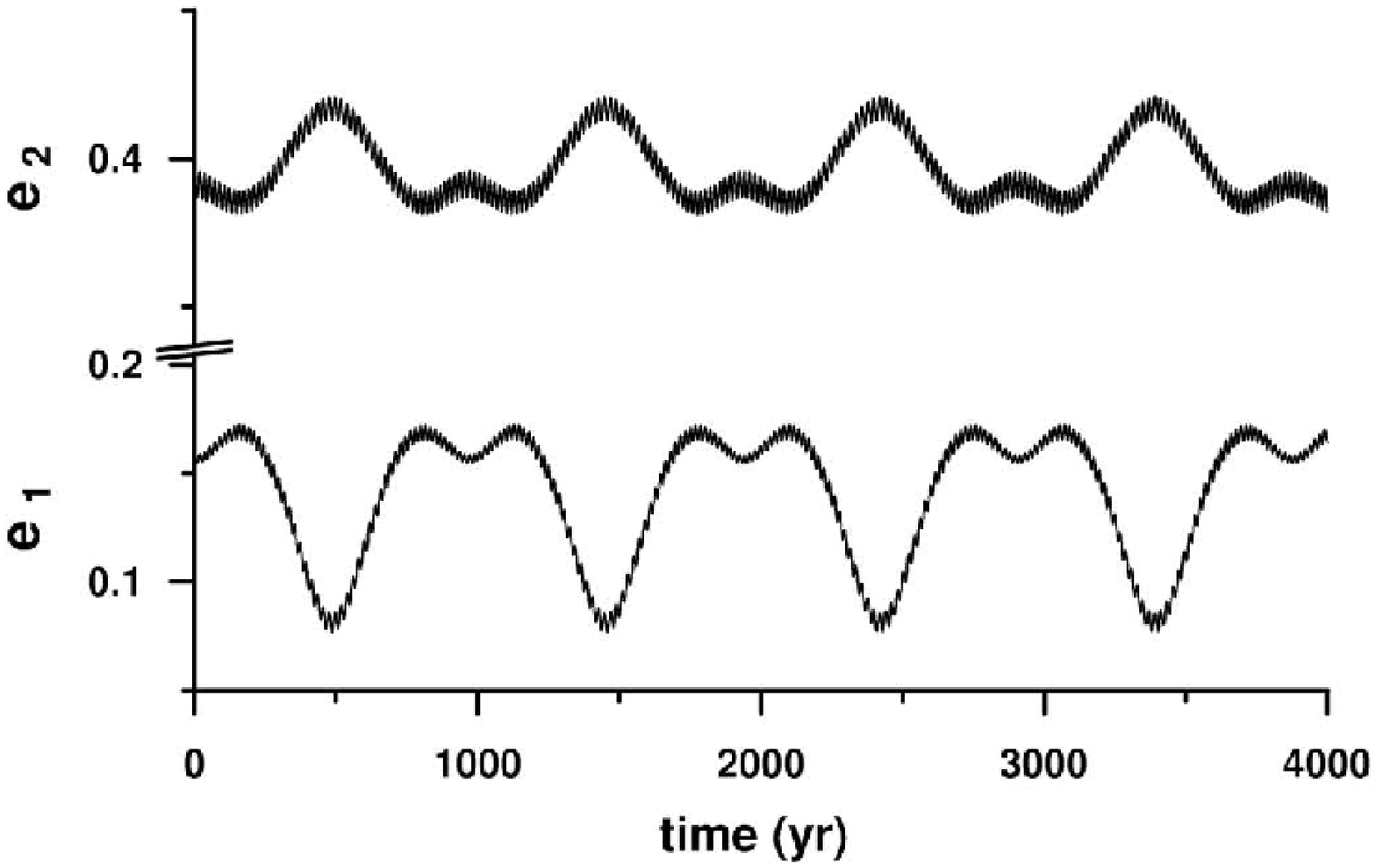}
  \caption{Numerical simulation of two planets displaying large-amplitude librations around a $(0,0)$-ACR, with masses and initial conditions drawn from Fit B of Ferraz-Mello et al. (2005) for the HD\,82943 system. Plot shows the resonant evolution of the semimajor axes (left panel) and eccentricities (right panel), averaged over short-term oscillations. It is worth comparing the resonant oscillation timescale (left panel) to the secular one (right panel). }
  \label{fm1}
\end{figure}

The invariance of ${\mathcal AM}$ and ${\mathcal K}$ has important consequences for the orbital evolution of the system. It indicates that, after the averaging process, from the set of the variables $(a_1,a_2,e_1,e_2)$, only two are independent, and the planar resonant problem has two degrees of freedom. Each degree of freedom is characterized by its proper mode of motion. Note that ${\mathcal K}$ depends only upon masses and semimajor axes; therefore, the invariance of the spacing parameter ${\mathcal K}$ defines a coupling of the semimajor axes: they oscillate with opposite phases and with amplitudes that are inversely proportional to the planetary mass. We say that the semimajor axes have no secular variation and their evolution occurs in timescales associated with the \textit{resonant mode} of motion. This effect can be seen in Figure \ref{fm1} (left panel).

On the other hand, the integral of total angular momentum ${\mathcal AM}$ causes a coupling of the planetary eccentricities, in such a way that, when one eccentricity grows, the other decreases, as seen in Figure \ref{fm1} (right panel). The oscillation generally occurs in timescales much longer when compared to those of the resonant mode. This mode is defined as \textit{secular mode} of motion of the system and is associated with the secular angle $\Delta \varpi$.  ${\mathcal AM}$ is a function of both semimajor axes and eccentricities, that implies that both modes of motion, resonant and secular, are present in eccentricity variations. In the example shown on the right panel in Figure \ref{fm1}, the amplitudes of the resonant component in the eccentricity variations are much smaller (and frequencies are much higher) than those of the secular mode. However, as will be shown below, for some initial conditions, both amplitudes and frequencies can be comparable.

\section{Stationary orbits of the averaged 2/1 resonant problem }\label{sec:2}

\begin{figure}[!ht]
  \centering
  \includegraphics[width=.45\textwidth]{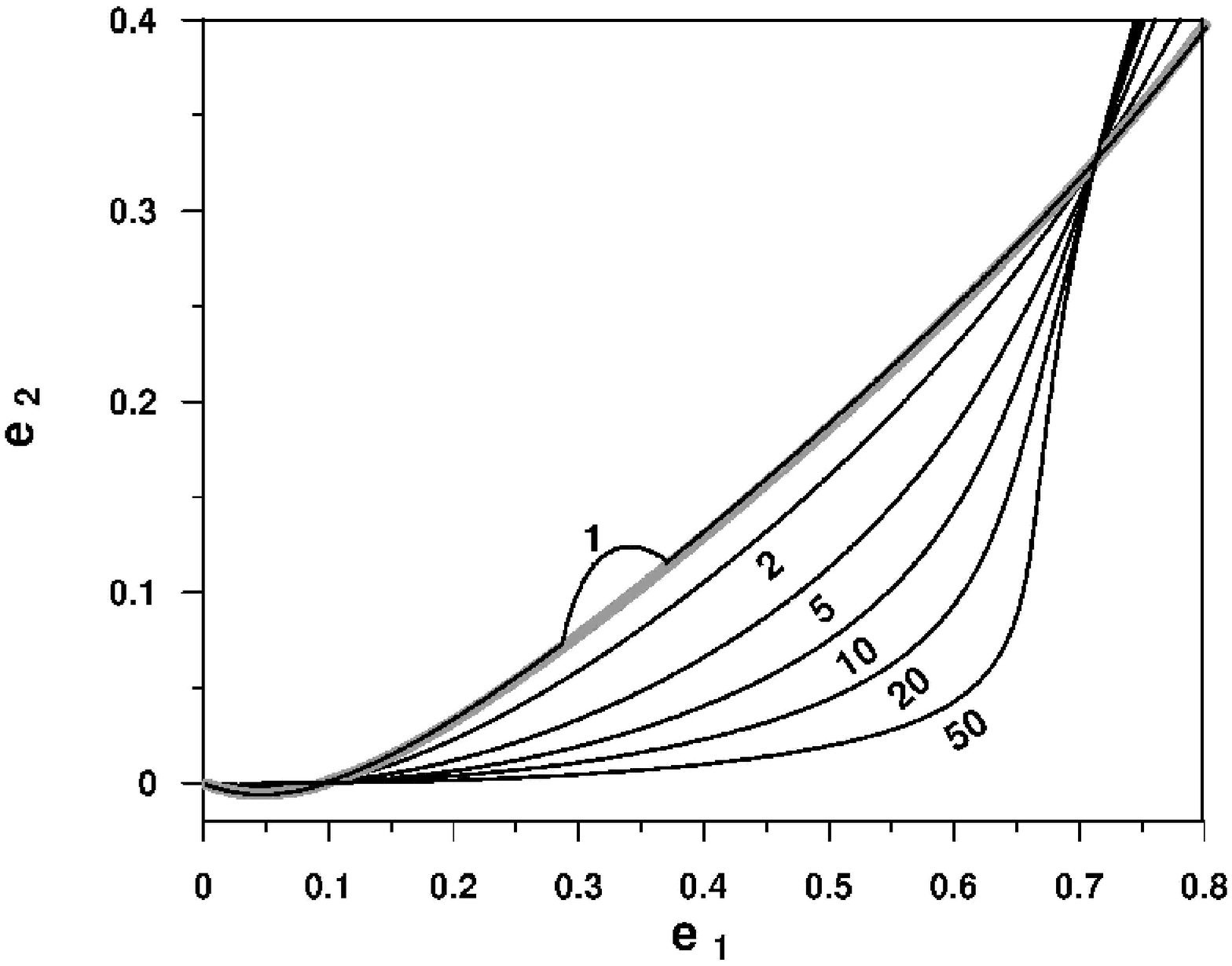}~\hfill
  \includegraphics[width=.45\textwidth]{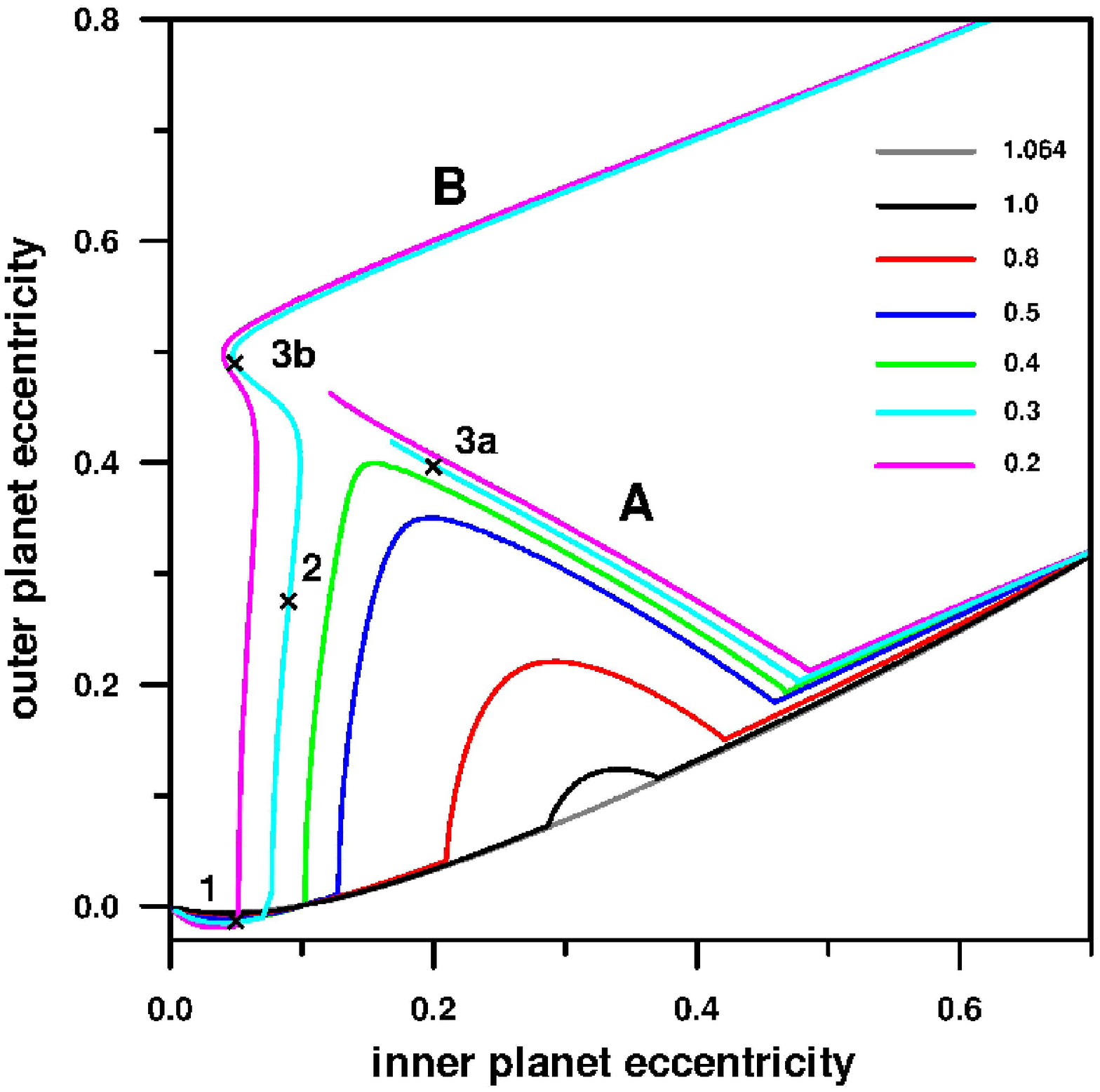}
  \caption{The 2/1 resonance stable apsidal co-rotation solutions parameterized by the mass ratio $m_2/m_1$. The cases with $m_2/m_1 \ge 1$ are shown on the left panel, while with $m_2/m_1 \le 1$ on the right panel. The mass ratio $1.064$ corresponding to the HD\,82943 system is illustrated by a gray curve.}
  \label{f1}
\end{figure}

The Hamiltonian function given in Eq.(\ref{eq1}) is very complicated, even in the planar case, and generally possesses several extrema, which define stationary solutions of the averaged problem. These solutions are often referred to as {\it Apsidal Corotation Resonances} (ACR). To obtain these special solutions over a large domain of the parameters of the problem, we employ the geometrical method presented in Michtchenko et al. (2006b), where each equilibrium solution is identified looking at local maxima of the Hamiltonian function, for given values of the total angular momentum ${\mathcal AM}$ and the spacial parameter ${\mathcal K}$. The location of the global maxima of the 2/1 resonance Hamiltonian calculated for all possible values of ${\mathcal AM}$ and a fixed ${\mathcal K}$, are shown on the plane ($e_1$, $e_2$) in Figure \ref{f1}, in the form of families parameterized by the mass ratio $m_2/m_1$.

Usually, ACR are classified in two types: symmetric and asymmetric solutions (e.g. Beaug\'e et al. 2003, Lee 2004, Voyatzis and Hadjidemetriou 2005). The symmetric solutions are characterized by stationary values of both critical angles (\ref{eq2}) at $0$ or $180^\circ$. The symmetric ACR--solutions evolve monotonically with the increasing eccentricities of the planets. For instance, for $m_2/m_1 \ge 1.015$ (Ferraz-Mello et al. 2003), all ACR families are symmetric; some are plotted on the left graph in Figure \ref{f1}.

For mass ratios smaller than this critical value, the smooth evolution along a symmetric family is interrupted by a sudden increase of the outer planet's eccentricity (the case of the family with $m_2/m_1=1$ on the left graph). The corresponding symmetric solutions become unstable and the stationary systems evolve now along an asymmetric segment of the corresponding family. Asymmetric ACR are characterized by stationary values of both critical angles different from zero or $180^\circ$. The families shown on the right graph in Figure \ref{f1} exhibit this feature.

A second critical value of the mass ratio occurs at $\simeq 0.36$ and is illustrated by cyan and magenta curves in Figure \ref{f1} (right panel). Below this limit, asymmetric ACR--families split into two disconnected branches on the ($e_1$,$e_2$)--plane, marked by labels {\bf A} and {\bf B}. In these cases, there may exist two distinct stationary configurations described by different sets of orbital elements, but leading to the same value of total angular momentum.

\begin{figure}[!ht]
  \centering
  \includegraphics[width=.45\textwidth]{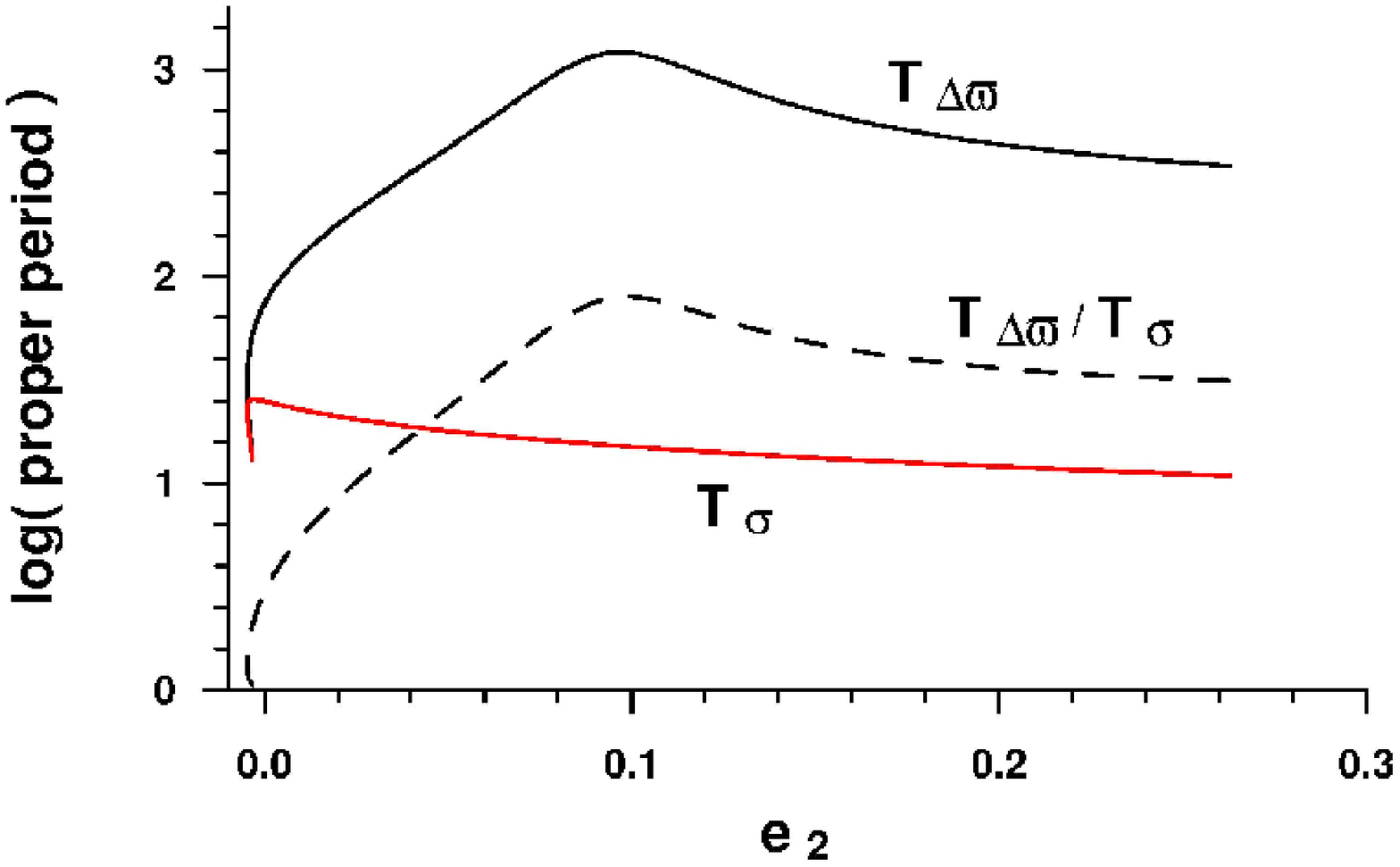}~\hfill
  \includegraphics[width=.45\textwidth]{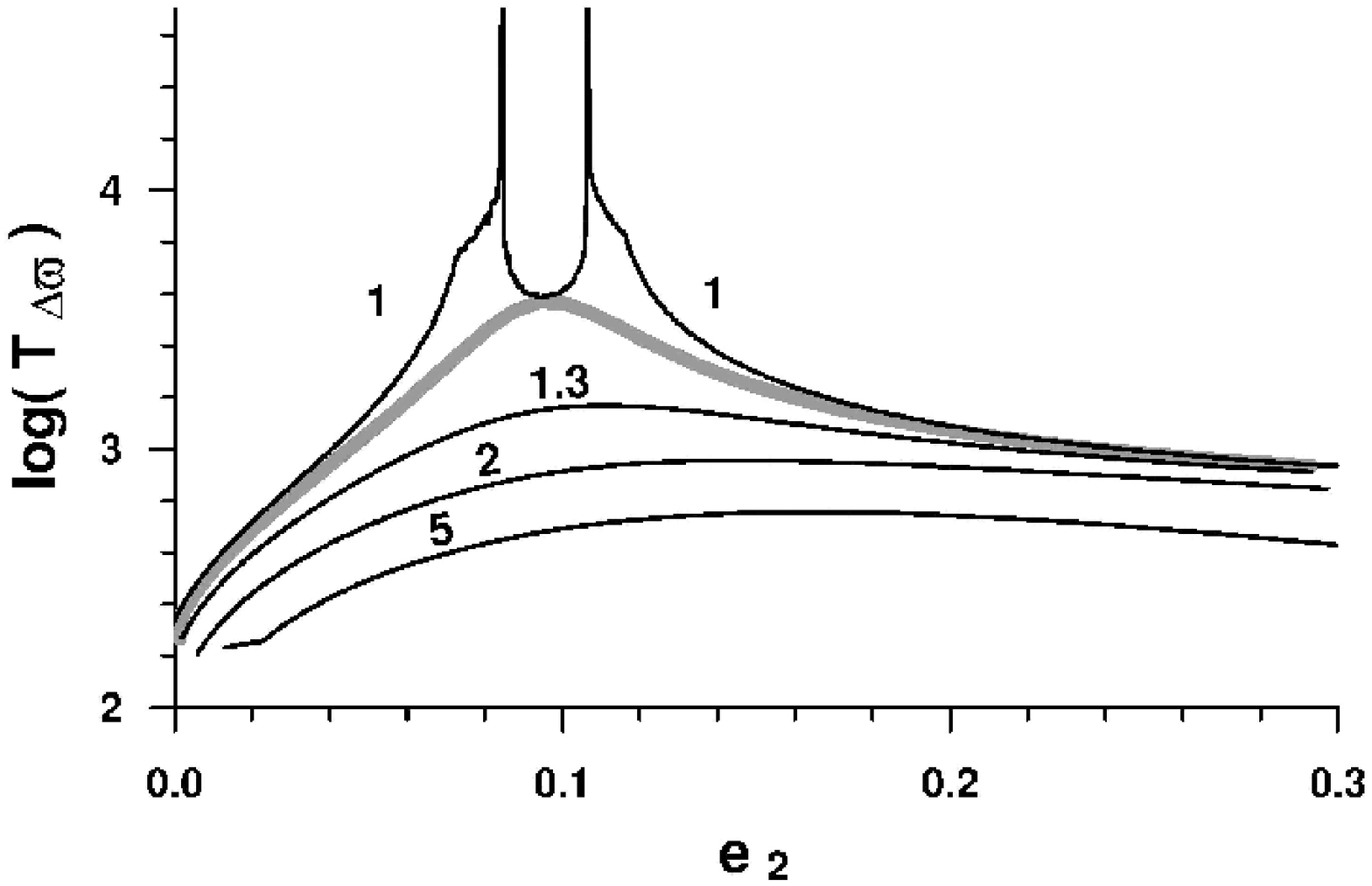}
  \caption{Left: Evolution of the period $T_{\sigma}$ of the resonant mode of motion and the period $T_{\Delta \varpi}$ of the secular mode along the ACR family parameterized by $m_2/m_1=1.064$, with the individual planet masses of the HD\,82943 system. The ratio $T_{\Delta \varpi}/T_{\sigma}$ is shown by dashed line. Right: Evolution of the secular period $T_{\Delta \varpi}$ along several ACR families. The values of the mass ratio are indicated by numbers.}
  \label{f2}
\end{figure}

We analyze the evolution of the proper periods (inverse of the proper frequencies) along each ACR--family, employing a method of dynamic power spectra (Michtchenko et al. 2006a). The left graph in Figure \ref{f2} shows, in logarithmic scale, the period of the resonant proper mode of motion ($T_\sigma$) and the period of the secular mode ($T_{\Delta\varpi}$), both obtained along the ACR-family parameterized by the mass ratio of the HD\,82943 system. The ratio between both periods, plotted by a dashed line on the same panel, indicates that the characteristic times of the resonant component of motion are generally shorter (at least in order $1$) than the secular ones. Both periods are comparable solely for very small eccentricities when, for some initial configurations of the planets, they can be commensurable originating a complex structure of \textit{secondary resonances} inside the primary 2/1 resonance. %Remember that secondary resonances take place at initial conditions, for which two proper frequencies become commensurable.

The right graph in Figure \ref{f2} shows the secular period, in logarithmic scale, as a function of the outer planet eccentricity obtained along several ACR--families with different mass ratios. For large values of $m_2/m_1$, there is a little change in the period with a maximum value at $e_2$ roughly between $0.1$ and $0.2$. However, for decreasing values of $m_2/m_1$, the maximum becomes more pronounced until it reaches a singularity near $e_2 \simeq 0.1$, for the mass ratio $m_2/m_1 \cong 0.015$. In Figure \ref{f2} (right panel) we show discontinuities  which appear along the ACR--family parameterized by $m_2/m_1=1$. A comparison with the plot of Figure \ref{f1} shows that these discontinuities are associated to the bifurcation of the symmetric ACR leading to the origin of the asymmetric stable solutions. The existence of the singularities indicate that the difference between symmetric and asymmetric ACR is not restricted to the equilibrium values of the angles, but constitutes qualitatively distinct solutions, each belonging to separate regions of the phase space.
\begin{figure}[!ht]
  \centering
  \includegraphics[width=.65\textwidth]{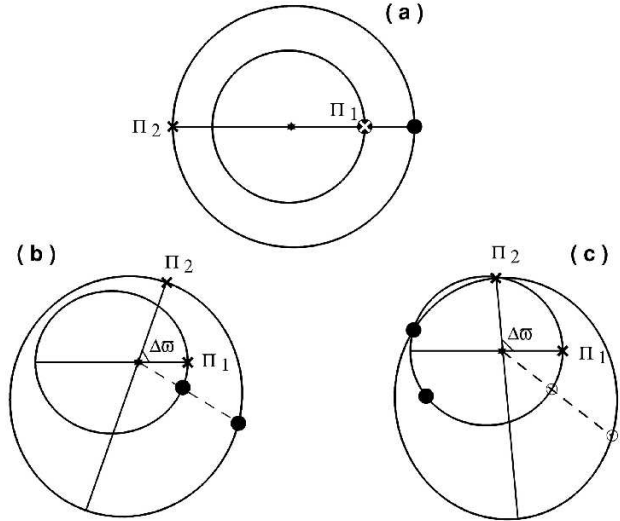}
  \caption{Examples of stable symmetric \textbf{(a)} and asymmetric \textbf{(b)} and \textbf{(c)} stationary configurations of two planets in the 2/1 resonance, with the mass ratio $m_2/m_1=0.3$. The continuous lines are apsidal lines, while the dashed lines are conjunction lines (in the case \textbf{(a)} two lines coincide). The full circles show the positions of the planets on the orbits at the instant of the closest approach between them. }
  \label{f3}
\end{figure}

\section{Geometry of the stationary orbits}\label{sec:3}

Figure \ref{f3} shows three different configurations of the ACR--solutions of the 2/1 MMR, obtained for $m_2/m_1=0.3$. The case \textbf{(a)} illustrates a symmetric solution with anti-aligned pericentre lines. In this case, the closest approach between two planets occurs when the planets are in conjunction: the inner planet in the pericentre and the outer planet in the apocentre of their orbits ($\Delta\varpi =180^\circ$).

Two asymmetric ACR-solutions are shown in Figure \ref{f3}\,(b) and (c). Both configurations are characterized by the same set of the constants ${\mathcal AM}$ and ${\mathcal K}$, but the solution \textbf{(b)} belongs to the branch {\bf A} of the ACR--family, while the solution \textbf{(c)} corresponds to the branch {\bf B} in Figure \ref{f1} (right panel). The secular angle $\Delta\varpi$ defines the relative orientation of the pericentre lines of two orbits. The positions of the planets on their orbits are strongly tied by the 2/1 resonance relationship between the planetary mean anomalies which are related to the critical angles through $M_1=\sigma_1+(p+q)\,Q$ and $M_2=\sigma_2+p\,Q$,where $Q$ is the synodic angle and $p=q=1$ (Beaug\'e and Michtchenko 2003). Full circles in Figure \ref{f3} show positions of the planets, for which their mutual distance is minimal.

In the configuration \textbf{(b)}, when the orbits are not intersecting, the closest approaches occur when the planets are in conjunction, forming a co-linear configuration with the central star. It is interesting to observe that the planetary conjunctions avoid positions between the two pericentra. As a consequence, conjunctions always occur when two planets are far away from positions where the two orbits are very close one to another; this feature can be seen as a protection mechanism due to the 2/1 mean-motion resonance.

In the case \textbf{(c)}, two orbits are intersecting and the mutual distance is not more minimal when the planets (open circles) are in conjunction. The closest approaches occur when the central star and the two planets (full circles) are in a triangle configuration and the outer planet is localized close to the intersection of the two orbits. In this case, again, the protection mechanism prevents very close approaches between two planets (see also Lee et al. 2006, for similar analysis).

\section{Two regimes of resonant behavior: interior and exterior resonances}\label{sec:4}

In the previous sections, we have seen that the resonant behavior is actually composed of two proper modes of motion, the resonant and secular ones. It is generally characterized by the libration of one angle and the circulation of the other. We refer to the librating angle as the \textit{resonant angle} and to the oscillation/circulation angle as the \textit{secular angle}. The secular theories provide us with the secular angle: it is defined by the difference in longitudes of pericentre $\Delta\varpi = \varpi_2-\varpi_1$ or, from our definitions (\ref{eq2}), as $\Delta\varpi = \sigma_1-\sigma_2$. The last expression shows that only one of two critical angles defined in Eq. \ref{eq2} is independent and is a truly resonant angular variable of the problem (the other independent angular variable is $\Delta\varpi$). The question which rises now is which one of two critical angles can be chosen as a resonant variable.

When the outer body is more massive ($m_2/m_1>1.015$), the situation is similar to that of an asteroid evolving in a resonance with Jupiter, when the librating angle is $\sigma_1$. In this case, all ACR-solutions are symmetric. Thus, we choose the critical angle $\sigma_1$ as a resonant angular variable of the problem and adopt the name {\it interior resonance} for this regime of motion.

For the planets with the mass ratios $m_2/m_1<1.015$, both kinds of ACR-solutions are possible, symmetric and asymmetric ones. If $m_2/m_1<<1$, which is the case of a Kuiper belt object in resonance with Neptune, the asymmetric solutions are dominating. Therefore, in the case of asymmetric ACR, we choose the critical angle $\sigma_2$ as a resonant angular variable of the problem. We refer to the regime of motion characterized by asymmetric ACR as {\it exterior resonance}.

The behavior of the secular angle $\Delta\varpi$ has also some special characteristics. In both interior and exterior resonances, it may be either a circulation or an oscillation. However, the oscillation in this case should be differentiated from the regime of motion defined as a libration in Celestial Mechanics. Indeed, the motions are akin to a family of concentric curves around one center displaced from the origin; the curves which are close to the center, do not enclose the origin -- they correspond to oscillations, while the outer curves enclose the origin and correspond to circulations. The separation between them is not a dynamical separatrix, but just one curve passing through the origin. The whole set of curves forms a homeomorphic family of solutions and the distinction between oscillations and circulations in this case is merely kinematical. To stress this behavior we use the word {\it oscillatiory/circulatory}, when describing this regime of motion.

\section{Dynamics around the ACR solutions: Interior resonance}\label{sec:5}

\begin{figure}[!ht]
  \centering
  \includegraphics[width=.65\textwidth]{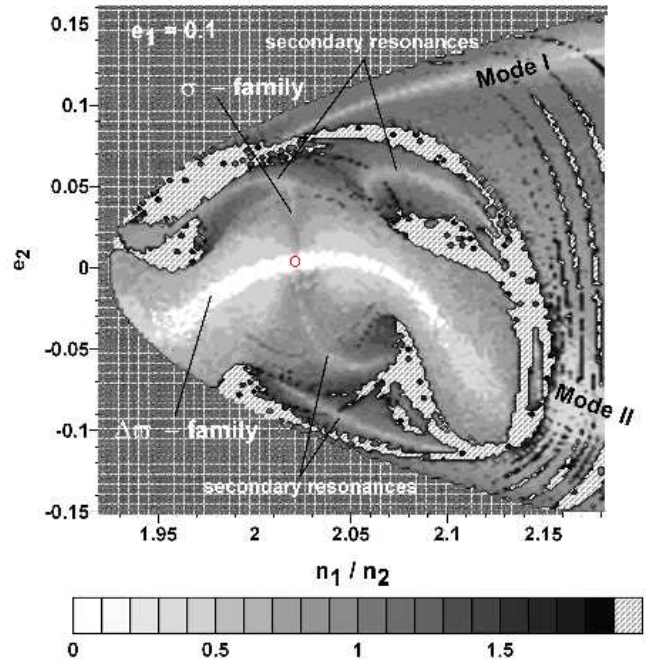}
  \caption{Dynamical map of the domain around the stable symmetric $(0,0)$--ACR (red dot) with $e_1=0.1$ and $m_2/m_1=1.064$. The initial value of $\sigma_1$ is fixed at 0 and the initial value of $\Delta\varpi$ is fixed at $0$ (positive values on the $e_2$-axis) or $180^\circ$ (negative values on the $e_2$-axis).}
  \label{fig:5}
\end{figure}

Independent of the mass ratio, the dynamics associated to all symmetric ACR is qualitatively similar, showing interior resonances. To portray the phase space of the interior resonance, we introduce a symmetric representative plane ($n_1/n_2$, $e_2$), with the resonant angle $\sigma_1$ fixed at 0 and the secular angle $\Delta\varpi$ at $0$ (positive values on the $e_2$-axis) or $180^\circ$ (negative values on the $e_2$-axis). A typical example of the dynamical map of the phase space around one symmetric ACR is shown in Figure \ref{fig:5}.

The map contains one center, which represents the ACR solution; its position shown by a red circle. Around the center, we find a region of quasi-periodic motion, coded in gray scale on the dynamical maps. Light gray color indicates regular motion, while darker tones correspond to increasingly chaotic motions.

The quasi-periodic behavior is the composition of two independent modes of motion: the resonant (with the frequency $f_{\sigma}$) and the secular (with the frequency $f_{\Delta\varpi}$) ones. The first one is associated to the resonant angle $\sigma_1$, while the second one to the difference in longitudes of pericentre, $\Delta\varpi$. Thus, any regular solution will be given by a combination of two periodic terms and their harmonics, each with a given amplitude and phase angle. Generally, both modes are well separated in the frequency space, with $f_\sigma$ much higher than the secular frequency $f_{\Delta\varpi}$ (see Figure \ref{f2} left panel). However, at small eccentricities, both frequencies may have same order, and it is possible to find initial conditions corresponding to low-order commensurabilities between them. They give origin to secondary resonances inside the primary 2/1 resonance, whose locations on the representative plane are indicated in Figure \ref{fig:5}.

For some initial conditions, the amplitude (not the frequency) of one mode tends to zero, and the solution changes from quasi-periodic to periodic. Since we have two independent frequencies, we can also have two independent families of periodic motion inside the 2/1 resonance, which we refer to as {\it $\sigma$-family} (the amplitude of the resonant mode is equal to zero) and {\it $\Delta\varpi$--family} (the amplitude of the secular mode is equal to zero). On the dynamical map shown in Figure \ref{fig:5}, the periodic families appear as continuous narrow white strips inside the gray tone domains of quasi-periodic motion. By definition, the intersection of two families gives us the position of the central ACR solution.

Apart from the characteristic frequencies, the resonant and secular modes also differ in dynamical behavior. The resonant mode is always librating and the passage to circulation must occur at a separatrix, which gives origin to domains of highly chaotic motion, with possible disruption of the system. At variance, the secular mode is generally oscillating/circulating; only in the domains of very high eccentricities it is possible to find initial conditions which lead to a true secular resonance inside the 2/1 MMR (see Section \ref{sec:6-2}).

The domains of chaotic motion (and large instabilities) are always present on dynamical maps of the 2/1 resonance. The chaotic behavior is associated with the existence of separatrices between: {\it (i)} the 2/1 resonance region and the regions of near-resonant and purely secular motion; {\it (ii)} the regions of qualitatively distinct regimes of motion inside the 2/1 mean-motion resonance, and {\it (iii)} secondary resonances inside the primary 2/1 resonance. The domains of highly chaotic motion are always shown in dark tones on the dynamical maps, while the hatched regions are regions of large-scale instabilities followed by disruption of the system within the time interval of each simulation (130,000 years). Finally, the domains of forbidden motion are filled by dark gray color in Figure \ref{fig:5}.
\begin{figure}[!ht]
  \centering
  \includegraphics[width=1.\textwidth]{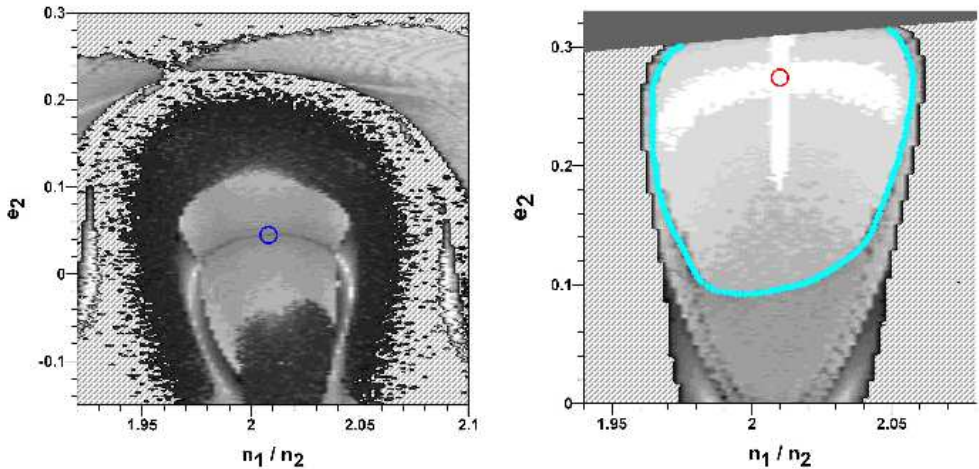}
  \caption{Left panel: Dynamical map on the representative plane obtained for $\sigma_1$ fixed at 0 and $\Delta\varpi$ at $0$ (positive values of the $e_2$-axis) or $180^\circ$ (negative values of the $e_2$-axis). The position of the unstable ACR is shown by a blue circle. Right panel: Dynamical map on the asymmetric representative plane obtained with initial values of the angular variables equal to those an asymmetric ACR: $\sigma_1=45.4^\circ$ and $\sigma_2=-68.4^\circ$, for $m_2/m_1=0.3$. The position of the stable ACR is shown by a red circle.}
  \label{fig:6}
\end{figure}

\section{Dynamics around the ACR solutions: Exterior resonance}\label{sec:6}

The dynamics associated to all asymmetric ACR is characteristic of exterior resonances. Asymmetric stationary solutions arise for mass ratios $m_2/m_1$ lesser than 1.015 (see Figure \ref{f1}, right panel). Moreover, for $m_2/m_1<0.36$ and at high eccentricities, each asymmetric family bifurcates into two disconnected branches on the ($e_1$, $e_2$)--plane, referred to as branches {\bf A} and {\bf B} in our work. The ramification of the asymmetric families indicates the advent of the secular resonance inside the mean-motion resonance. As a consequence, the phase space of the exterior resonance shows a very complex picture marked by the presence of several distinct regimes of resonant and non-resonant motion, crossed by families of periodic orbits and separated by chaotic zones.

\subsection{Dynamical map around the one-branch asymmetric ACR}\label{sec:6-1}

One-branch asymmetric solutions are characteristic for the mass ratio values in the range from 0.36 to 1.015. The dynamics around a single asymmetric ACR is illustrated in Figure \ref{fig:6} on the two representative planes: the left graph is a plane analogous to one introduced for the interior resonance and shown in Figure \ref{fig:5}; the right graph is a plane defined by the initial values of the resonant angles fixed exactly at those of the stable asymmetric stationary solutions.

The plane defined by symmetric resonant angles does not allow us to fully describe the dynamics of the exterior resonance. The basic properties of the resonant phase space, such as the existence of the intersecting families of periodic orbits and, thus, of the stable centers, are not seen on the resulting dynamical map. However, the analysis of this map allows us to detect the co-existence of the two regimes of motions named as interior resonance and exterior resonance for the same set of the constants $m_2/m_1$, ${\mathcal AM}$ and ${\mathcal K}$. Note the $\Delta \varpi$-families of symmetric periodic solutions (with $\Delta\varpi=0$ or $180^\circ$), which appear as narrow light strips inside the chaotic regions on the map in Figure \ref{fig:6}, left panel.

To better represent the dynamics in the neighborhood of the asymmetric centers, we re-calculated the dynamical map for the same sets of constants of motion, but with the initial values of the resonant angles fixed exactly at those of the stable asymmetric stationary solutions (right graph in Figure \ref{fig:6}). Now we can observe the existence of the stable center at the intersection of the two periodic families (white strips), $\sigma$ and $\Delta\varpi$. The stable domain surrounding the center (in light tones), is composed of quasi-periodic motions with two independent modes: resonant libration around the $\sigma$--family and secular oscillation/circulation around the $\Delta\varpi$--family. Generally, the global dynamics of the single exterior resonance is similar to the structure of the interior resonance presented in the previous section. However, there are two main differences: {\it (i)} the librating angle is $\sigma_2$ and the oscillating/circulating angle is $\sigma_1$ (and $\Delta\varpi$), and {\it (ii)} the angles oscillate around values that are not simply $0$ or $180^\circ$.
\begin{figure}[!ht]
  \centering
  \includegraphics[width=1.\textwidth]{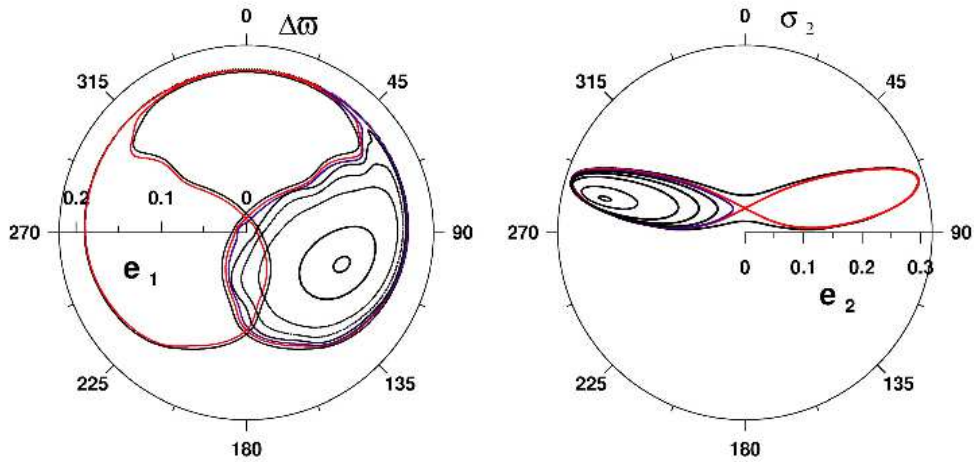}
  \caption{Smoothed planetary paths around one-branch asymmetric ACR in the ($e_1$, $\Delta\varpi$) (left panel) and ($e_2$, $\sigma_2$) (right panel) polar coordinates calculated with $e_1=0.09$ and $m_2/m_1=0.3$ and the same energy. The smoothing has been done using a low-pass filter.}
  \label{fig:7}
\end{figure}

The planetary motions around a stable asymmetric ACR are shown in Figure \ref{fig:7}, in the polar coordinates ($e_1$,$\Delta\varpi$) on the left graph and in the polar coordinates ($e_2$, $\sigma_2$) on the right graph. Only the secular components are presented; the fast resonant oscillations were eliminated using a low-pass filter. All planetary paths enclose the asymmetric ACR whose coordinates are $e_1=0.13$, $e_2=0.26$, $\sigma_2=284^\circ$ and $\Delta\varpi=108^\circ$. The orbits close to the ACR are presented by black curves: they show oscillations of both the resonant angle $\sigma_2$ and the secular angle $\Delta\varpi$. For initial conditions far away from the stable center, the amplitudes of the oscillations increase and the angle $\Delta\varpi$ starts to circulate, as shown by blue curve in Figure \ref{fig:7}, left panel. In this case, the curve separating the oscillatory and circulatory solutions represents a solution passing through the origin $e_1=0$; we say that the secular angle $\Delta\varpi$ (and $\sigma_1$) is in the oscillatory/circulatory regime of motion.

At variance, the red curve in Figure \ref{fig:7}, right panel, separating the oscillations of the resonant angle $\sigma_2$ around the asymmetric ACR from the oscillations around zero, is formed by solutions asymptotic to the unstable symmetric ACR. Hence this curve is a true separatrix and the oscillations of $\sigma_2$ around the stable asymmetric ACR are true librations.

Outside the separatrix, the resulting orbits (as an example, the external black curve in Figure \ref{fig:7}) are  \textit{horseshoe-like orbits} encompassing both asymmetric centers and appearing to oscillate around zero (analogous of orbits known from the dynamics of the asteroidal 1:1 mean-motion resonance with Jupiter). On the dynamical map on the right panel in Figure \ref{fig:6}, the domain of the horseshoe orbits is separated from the central resonance region by the blue curve; this curve is the locus of the initial conditions corresponding to the true separatrix corresponding to the red color path in Figure \ref{fig:7}. The detailed analysis of the dynamics in its close vicinity reveals that the transition across the blue curve is topologically discontinuous, but the associated large-scale instabilities followed by disruption of the system are not observed on the dynamical map.

\subsection{Dynamical maps around the two-branch asymmetric ACR}\label{sec:6-2}
\begin{figure}[!ht]
  \centering
  \includegraphics[width=.65\textwidth]{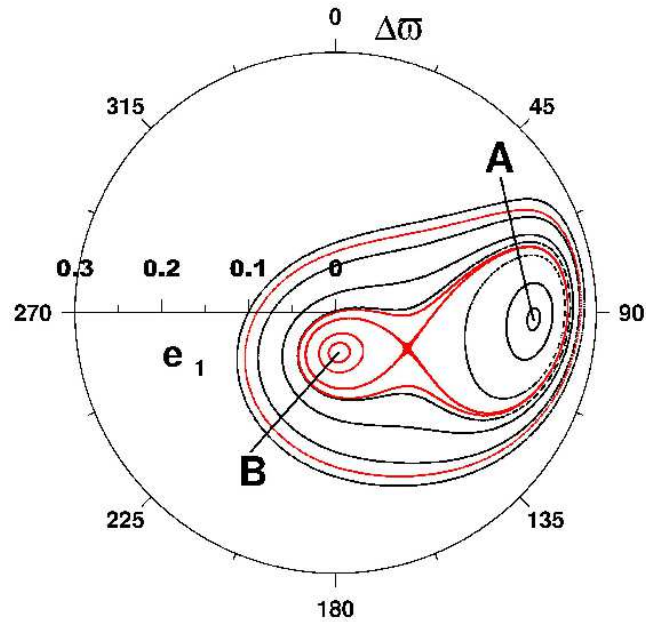}
  \caption{Smoothed planetary paths around two-branch asymmetric ACR in the ($e_1$, $\Delta\varpi$) polar coordinates calculated with $e_1=0.09$ and $m_2/m_1=0.3$ and the same energy. The smoothing has been done using a low-pass filter.}
  \label{fig:9}
\end{figure}

Two-branch asymmetric solutions are characteristic for mass ratio values lesser than 0.36. In this case, there exist two distinct stationary configurations described by different sets of orbital elements, but leading to the same value of total angular momentum. To better illustrate the bifurcation phenomenon, we analyze and compare planetary paths in the vicinity of the two-branch ACR: one belongs to the branch \textbf{A} and other to the branch \textbf{B}. This is done in Figure \ref{fig:9}, where, in order to simplify the visualization, we concentrate solely on the secular behavior of the system. The paths of both planets were smoothed using low-pass filtering (resonant oscillations were eliminated) and were plotted in polar coordinates ($e_1$, $\Delta\varpi$).

Two centers are now clearly observed in Figure \ref{fig:9}, instead of one center as in the case of the one-branch asymmetric ACR (see Figure \ref{fig:7}). The black curves surround the center associated to the branch {\bf A} of the periodic solutions, while the red curves surround the center on the branch {\bf B}. Among these paths, we find an infinite-period separatrix formed by solutions asymptotic to an unstable ACR (saddle point). Outside the separatrix, all paths follow similar patterns and are structurally stable. The existence of the separatrix, not present in the case of a single ACR, implies that there are motions near one of the two-branch asymmetric ACR that actually correspond to a true libration around the ACR and no longer a simple circulation/oscillation. This asymmetric ACR belongs to the branch {\bf B}. In its neighborhood therefore both $\Delta\varpi$ and $\sigma_2$ show true librations around the equilibrium values. In other words, the two-branch asymmetric ACR generates, inside the 2/1 mean-motion resonance, a {\it true resonance} of the secular angle $\Delta\varpi=\sigma_1-\sigma_2$. This novel behavior is absent in the cases of symmetric and one-branch asymmetric ACR (except for very-high, close to 1, eccentricities of the inner planet).
\begin{figure}[!ht]
  \centering
  \includegraphics[width=1.0\textwidth]{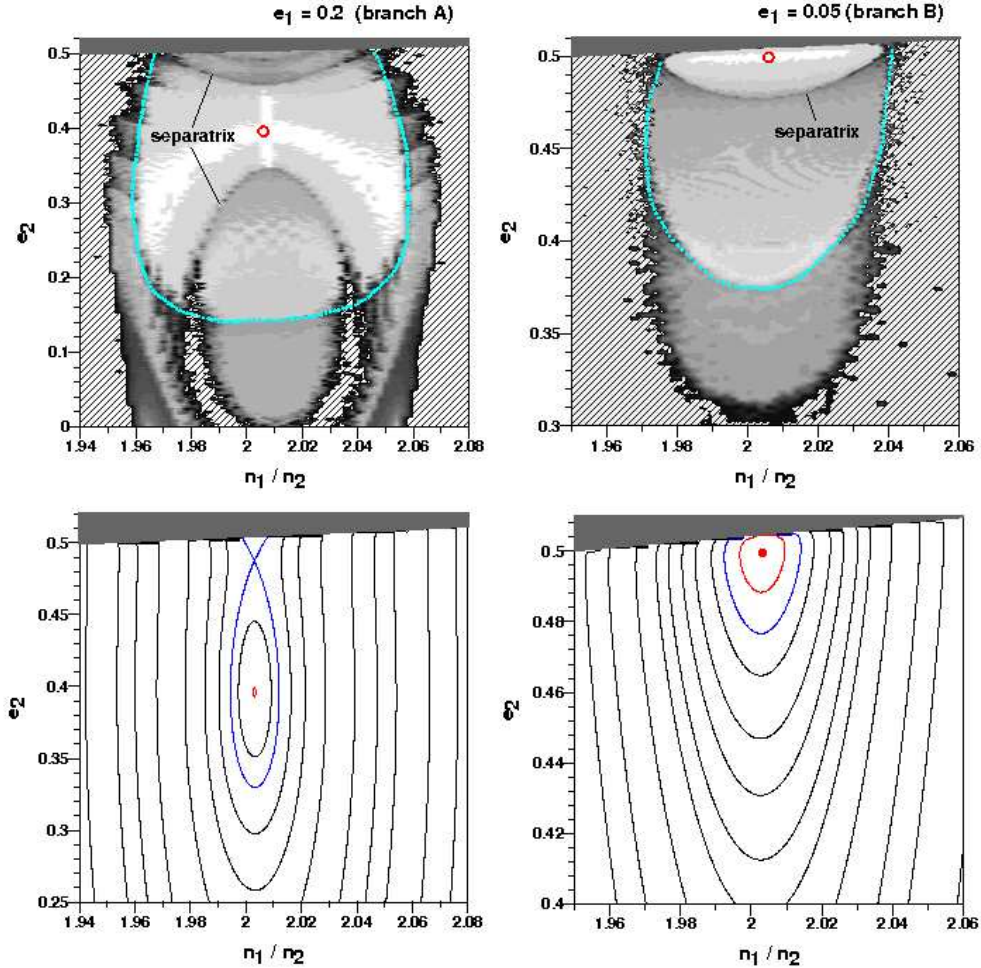}
  \caption{Top: Dynamical maps of the two possible asymmetric solutions: one with $\sigma_1=32.2^\circ$ and $\sigma_2=-62.4^\circ$ belongs to the branch {\bf A} (left panel), and other, with $\sigma_1=120.7^\circ$ and $\sigma_2=-69.1^\circ$, to the branch {\bf B} (right panel). The positions of the stable asymmetric centers on the corresponding dynamical maps are marked by red dots. Bottom: The levels of the resonant Hamiltonian around the branch {\bf A} ACR (left panel) and the branch {\bf B} ACR (right panel).}
  \label{fig:8}
\end{figure}

Using the same constants of motion, but two distinct sets of critical angles, we constructed the dynamical maps on the two asymmetric representative planes, shown in Figure \ref{fig:8} (top panels). The positions of the stable asymmetric centres on the corresponding dynamical maps are marked by red dots. The dynamical map of the branch {\bf A} solution (left-top panel) displays qualitative similarities with the one-branch ACR map shown in Figure \ref{fig:6}, right-top panel, mainly the existence of two families of periodic motion intersecting at the center and a large zone of the quasi-periodic motion inside the exterior resonance regime. The main difference is the existence of the separatrix whose location is indicated on the left-top graph in Figure \ref{fig:8}.

In order to better visualize a separatrix feature, we present in the same figure (bottom panels) the neighborhoods of the stable asymmetric ACR, plotting the energy levels of the resonant Hamiltonian. A bifurcation phenomenon is clearly observed in the behavior of the Hamiltonian around the branch \textbf{A} ACR on the left-bottom panel, where the energy level in blue color presents a saddle-type structure. The continuation of the same level in the domain of the branch \textbf{B} ACR appears on the right-bottom graph as a blue curve. The fixed point of the centre (red dot) on the left-bottom panel appears as the energy level in red color on the right-bottom graph; that is, the branch \textbf{B} stationary solution is located at the global maximum of the resonant Hamiltonian.

The region surrounding the branch \textbf{B} ACR (right-top panel) is a domain of true secular resonance, when the passage from oscillation to circulation of the angle $\Delta\varpi$ is chaotic. In this region, both the resonant angle $\sigma_2$ and the secular angle $\Delta\varpi$ librate around the corresponding asymmetric values (in the case shown in Figure \ref{fig:8}, they are $120.7^\circ$ and $189.1^\circ$, respectively). The equilibrium configurations of the planetary orbits are similar to those shown in Figure \ref{f3}: the closest approaches occur when the central star and the two planets form a triangle configuration. Our numerical simulations have shown that the planetary motions in this region are very stable, even for very high eccentricities of the orbits.

Finally, on both maps in Figure \ref{fig:8} there exist regions of the horseshoe-like orbits at initial conditions situated outside the cyan curves. The horseshoe orbits are not centered on any single ACR, but actually contain all of them including the single (or double) stable asymmetric ACR and the unstable symmetric ACR.

\acknowledgements This work has been supported by the Brazilian National Research Council - CNPq and the S\~ao Paulo State Science Foundation - FAPESP. The authors gratefully acknowledge the support of the Computation Centers of the University of S\~ao Paulo (LCCA-USP) and of the Astronomy Department of the IAG/USP, for the use of their facilities.

\end{document}